# Color Detection Using Chromophore-Nanotube Hybrid Devices

*Xinjian Zhou\*, Thomas Zifer, Bryan M. Wong, Karen L. Krafcik, François Léonard and*

*Andrew L. Vance*

Sandia National Laboratories, Livermore, CA, 94551

\*Corresponding author. E-mail: xinzhou@sandia.gov.

ABSTRACT

We present a nanoscale color detector based on a single-walled carbon nanotube functionalized with azobenzene chromophores, where the chromophores serve as photoabsorbers and the nanotube as the electronic read-out. By synthesizing chromophores with specific absorption windows in the visible spectrum and anchoring them to the nanotube surface, we demonstrate the controlled detection of visible light of low intensity in narrow ranges of wavelengths. Our measurements suggest that upon photoabsorption, the chromophores isomerize from the ground state *trans* configuration to the excited state *cis* configuration, accompanied by a large change in dipole moment, changing the electrostatic environment of the nanotube. All-electron *ab initio* calculations are used to study the chromophore-nanotube hybrids, and show that the chromophores bind strongly to the nanotubes without disturbing the electronic structure of either species. Calculated values of the dipole moments support the notion of dipole changes as the optical detection mechanism.

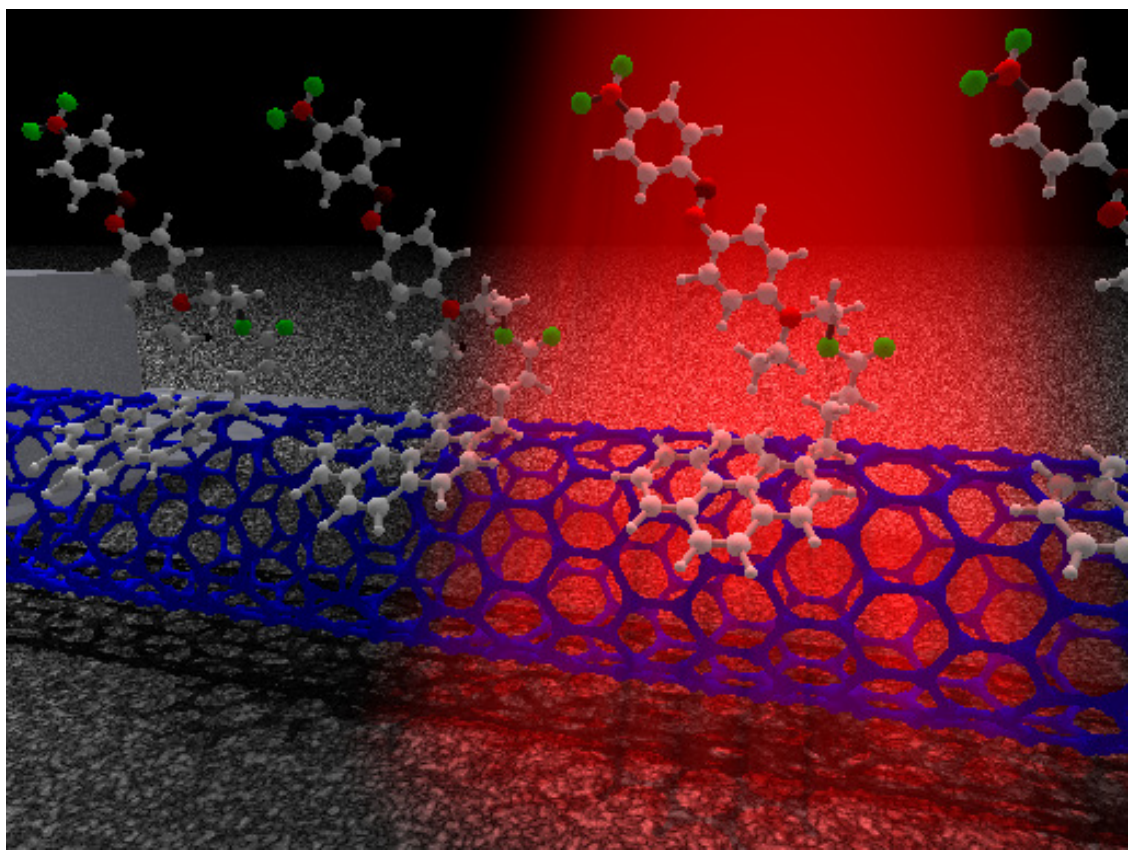

Phototransduction plays a central role in key natural processes such as vision and photosynthesis. For these processes, natural evolution has found exquisite solutions that optimize efficiency, compactness, and self-healing. While researchers have long tried to supplant nature's successes using solid-state approaches, our ability to convert optical radiation to electrical signals with nanoscale precision still remains in its infancy[1-4]. An alternative to solid-state approaches is one that more closely mimics natural processes by integrating chemical or biological materials with solid-state materials. An example of this approach is that of optical detection using functionalized single-walled carbon nanotubes (SWNTs). SWNTs' high sensitivity to local electrostatic environments[5-8] can be used to detect optically-induced changes in the light-absorbing molecules. Indeed, it has been shown that when networks of SWNTs are functionalized with porphyrin[9], charge transfer between the porphyrin and the nanotube upon illumination can be detected using a field-effect transistor (FET) geometry. Similarly, the transition of spiropyran to its charge-separated state upon UV photon absorption has been detected electrically using individual SWNTs[10].

Recently, a different class of nanotube functionalization that is sensitive to optical radiation has been proposed[11]. In this approach, one kind of azobenzene-based chromophore, Disperse Red 1 (DR1), was attached to SWNTs, and it was demonstrated that the conduction of such functionalized SWNTs can be modulated with UV light. This phenomenon is attributed to the photon-induced isomerization of the azobenzenes that changes the electrical dipole moment of the chromophore and thus modifies the electrostatic potential on the SWNTs. One question is whether these results can be extended to the visible spectrum, and in particular if different regions of the visible spectrum can be specifically targeted. In this study we address this question by assembling chromophore-nanotube FET hybrids where three azobenzene-based chromophores are designed to respond specifically to different regions of the visible spectrum. We report that these devices are highly sensitive to light, and that their response correlates with the absorption spectra of the attached chromophores. In addition, the magnitude of observed signals and their time dynamics are examined at various light intensities and wavelengths. Moreover, we present all-electron *ab initio* calculations of the interaction between the chromophores and SWNTs, including that of the dipole moments in the *trans* and *cis* configurations. These calculations allow us to support the notion that the optical-to-electrical signal transduction is controlled by a dipole change mechanism at the molecular level. This work demonstrates a way to make color photodetectors with nanoscale resolution and provides insight into the interactions between SWNTs and molecules in their close vicinity.

Carbon nanotube FETs used in this study were prepared with SWNTs purchased from CheapTubes. These tubes were grown using chemical-vapor deposition and have diameters in the range of 0.8-2 nm. SWNTs received in powdered form were ultrasonically dispersed in water with 2 wt% sodium dodecylbenzenesulfonate (SDBS) and purified via ultracentrifugation (36,000 rpm, 2 hours) to yield well-dispersed SWNTs[12]. Silicon wafers with a 100 nm thermal oxide capping layer were treated with (3-aminopropyl)triethoxysilane (APTES)[13] and incubated with the SWNT solution (The APTES layer provides positive surface charges to the silicon wafers to enhance the attachment of SWNTs, which are wrapped with negatively-charged surfactants). On top of these SWNTs, source-drain electrodes with a typical 2 μm gap were defined with photolithography, and either gold or palladium was evaporated to make electrical contacts. The density of nanotubes was controlled such that many source-drain gaps are each bridged by a single SWNT. Only semiconducting devices that can be completely shut off by the electrical potential on the highly-doped silicon backgate were selected for chromophore functionalization and testing.

Chromophores were prepared via dicyclohexyl carbodiimide (DCC) esterification/amidation similar to the chemistry reported by Simmons, et al. [11] for the preparation of anthracene-functionalized DR1. In this study, 1-pyrenebutyric acid was used instead of 9-anthracenecarboxylic acid, and, in addition to DR1, disperse orange 3 (DO3) and 4-(4-nitrophenyl)azophenol (NPAP) were functionalized with pyrenebutyric acid. A non-isomerizable control molecule was prepared by esterification of 1-pyrenebutyric acid with hexanol. The products are abbreviated as DR1-PB, DO3-PB, NPAP-PB and hexyl-PB. Structures are shown in Fig. 1b. Products were characterized by $^1$H-NMR spectroscopy, FTIR, mass spectrometry and UV/Vis spectrophotometry (see the Supporting Information for details).

In the final step of device fabrication, SWNT FETs were incubated with the chromophore solution for 1 minute and subsequently washed lightly with methanol. The chromophores selectively deposited onto SWNTs via strong non-covalent interactions between nanotube sidewalls and the pyrene linker group[14] of the chromophores.

Electrical-optical studies were performed on a modified Zeiss optical microscope. Light from a xenon arc lamp was fed into an Acton Spectropro-275 monochromator with a 600 g/mm grating; and the resulting monochromatic light with a 3 nm bandwidth was focused through a long working distance 50X objective to a spot of about 35 μm in diameter, small enough to illuminate only one SWNT FET device. The typical power density of illumination was 100 W/cm$^2$, and the wavelengths ranged from 400 nm to 700 nm. The current of SWNT FETs was measured with an Ithaco 1211 (DL Instruments) preamplifier under a 10 mV DC bias applied between the source and drain electrode.

Figure 1a shows the transfer characteristics of a DR1-PB functionalized device in the dark and when illuminated at different wavelengths. While the device was insensitive to 400 nm violet and 630 nm red light, 470 nm and 530 nm radiation shifted the threshold voltage to the positive side by about 1 volt. The transfer characteristics recovered fully once the illumination was turned off. Control experiments were carried out to narrow down the possible origins of the signal. First, bare SWNT FETs were tested and confirmed to be insensitive to visible light at this power level. Second, hexyl-PB molecules with their dipole-free alkyl chain were attached to the SWNT and these devices did not respond to visible light at a similar power density as the DR1-PB functionalized devices. Last, the signal disappeared when the 35 μm large light spot was moved slightly away from the devices, eliminating the influence from the silicon backgate as the origin of the signal.

To further our understanding of the system, we measured the threshold voltage shift as a function of wavelength between 400 nm to 650 nm. Data at different wavelengths were taken with a random sequence to minimize systematic drifting, and the conductance in the dark was checked after each exposure to ensure that the device recovered completely. The results of our measurements are plotted in Fig. 2. The magnitude of the threshold voltage shift at different wavelengths is similar to the absorption of DR1 molecules measured in 1,2-dichloroethane (DCE) at 1 mM concentration. The peak in the threshold voltage shift is broader than the absorption peak, and is shifted towards the red, an effect that could be attributed to changes in absorption generated by the local electrostatic environment. Indeed, this phenomenon has been seen in previous studies of the absorption properties of DR1 in different solvents[15]. Therefore, we are confident in linking the threshold voltage shift in the conduction of SWNT FETs to the photon absorption of DR1-PB molecules. This linkage is further strengthened, as shown in Fig. 2, when the threshold voltage shift of more devices functionalized with either DO3-PB or NPAP-PB molecules are shown to shift according to the absorption spectra of these two molecules. Thus, by choosing specific chromophores, we obtain optical detection in specifically targeted regions of the visible spectrum.

In order to gain more insight into the properties of the chromophore-nanotube hybrids, we performed all-electron *ab initio* calculations of the structural and electronic properties of SWNTs functionalized with DR1-PB, DO3-PB and NPAP-PB. It is well understood that when these chromophores absorb photons, the azobenzenes isomerize from the ground state *trans* configuration to the metastable excited state *cis* configuration, with a large change in dipole moment[16]. Thus we performed calculations for both the *cis* and *trans* configurations for the three molecules attached to SWNTs. First, geometry optimizations for all *cis* and *trans* chromophores were obtained from density functional theory (DFT) calculations using the Slater-Vosko-Wilk-Nusair local density functional[17] in conjunction with a 6-31G(d,p) Gaussian basis set. A (10,0) semiconducting SWNT was chosen due to its large enough size to approach those measured experimentally, but small enough to reduce computational costs. Calculations were performed using a one-dimensional supercell along the axis of the nanotube. Since the chromophore molecules are over 5 times longer than the (10,0) unit cell, a large supercell of 12.8 Å along the nanotube axis was chosen, which allows a separation distance of 3.6 Å between adjacent chromophores. Figure 3 shows the relaxed atomic structure for all six possible combinations of the three molecules in their *trans* and *cis* configurations. Based on these atomic structures, we found binding energies that vary between -0.19 eV and -0.24 eV. Charge transfer between the molecules and the SWNT was found to be 0.04-0.05 electrons from the molecule to the SWNT, and, more importantly, the amount of charge transfer varies little between the *trans* and *cis* conformations of the same molecule. These results for the binding energy and charge transfer are typical of DFT calculations within the Local Density Approximation (LDA), which are known to overestimate the binding energy in systems governed by van der Waals forces, such as those between the pyrene linker and the SWNT. Our results should thus be considered as upper bounds for the binding energy and charge transfer. We found that the electronic band structure of the SWNT is unaffected by the presence of the chromophores, and that the molecular energy levels of the chromophores are unchanged and dispersion-less.

From these results, we calculated the electric dipole moments of the chromophores as expectation values of the dipole operator using the Kohn-Sham density matrix. Since the charge transfer between the molecules and the SWNT is essentially the same for the *trans* and *cis* configurations, we calculated the dipole moments for the isolated molecules using the relaxed atomic positions from the molecule+SWNT system while enforcing charge-neutrality on the molecule. This approach allowed us to obtain values of the dipole moments that are independent of the origin of the coordinate system. The values of dipole moments were calculated to be 13.18 D (*trans*) and 11.60 D (*cis*) for DR1-PB, 10.81 D (*trans*) and 9.90 D (*cis*) for DO3-PB, and 8.27 D (*trans*) and 7.25 D (*cis*) for NPAP-PB. To understand how these molecular dipole shifts can lead to large changes in the threshold voltage shift of a SWNT FET, we calculated the electrostatic potential on a SWNT functionalized with a uniform density of 1 molecule/nm, with the dipole moments oriented perpendicular to the SWNT axis and forming a line. Treating the molecular dipoles as point dipoles, the electrostatic potential on a line on the nanotube surface closest to the line of dipoles is given by $V = \frac{\mu d}{4\pi\varepsilon_0} \sum_{i=-\infty}^{\infty} \left[ (h/\xi)^2 + i^2 \right]^{-3/2}$ where the index $i$ refers to the $i$th dipole along the SWNT, $h$ is the smallest distance between a dipole and the SWNT and $\xi$ is the distance separating the dipoles along the SWNT. This expression allows calculation of the electrostatic potential difference between the *trans* and *cis* configurations as $\Delta V = V_{trans} - V_{cis}$. To obtain $\Delta V$ for the three chromophores used in this study, we calculated $h$ from the "center of charge mass" for each of the six possible cases using the relaxed atomic positions of the molecule on the SWNT obtained from the *ab initio* calculations. From these values and the calculated dipole moments, we obtained $\Delta V_{DR1} = 0.14$ V, $\Delta V_{DO3} = 0.26$ V, and $\Delta V_{NPAP} = 0.1$ V. This can be related to the change in gate voltage using the ratio of capacitances for the gate and the molecules. The total capacitance for the gate is given by $C_{tot}^{-1} = C_g^{-1} + C_{NT}^{-1}$ where $C_g = 2\pi\varepsilon / \ln(4h/d) = 0.039 \; aF/nm$ is the gate capacitance for an oxide thickness $h = 100$ nm and a SWNT diameter $d = 1.5$ nm, and $C_{NT}$ is the intrinsic NT capacitance. For the molecules, the only contribution to the capacitance is $C_{NT}$ since the pyrene polarization is weak. $C_{NT}$ can be calculated from the expression $C_{NT} = ed\lambda / dE_F$ where $\lambda$ is the linear charge density on the SWNT and $E_F$ is the Fermi level. The dependence of $\lambda$ on the position of the Fermi level can be obtained by integrating the SWNT density of states times the Fermi distribution[18], giving $\lambda \approx 2e(3\pi\gamma a\beta)^{-1} \exp\left[-\beta(E_F - E_v)/kT\right]$. Here, $a$ = 0.142 nm is the C-C bond length, $\gamma = 2.5\ eV$ is the value of the tight-binding overlap integral, and $\beta \approx 0.7$ is a numerical factor[18]. Near the threshold voltage, the Fermi level is close to the valence band edge and we obtain $C_{NT} = 0.75\ aF/nm$ for a SWNT of bandgap 0.5 eV and diameter 1.5 nm. Finally, an electrostatic potential change $\Delta V$ due to a change in the molecular dipole moments translates into a gate voltage shift $\Delta V_g = (C_{NT}/C_{tot})\Delta V \approx 20\Delta V$. Thus, a relatively small change in the electrostatic potential of 0.1 V is amplified into a large 2 volt threshold voltage shift, as observed in our measurements.

We now turn to the kinetics of the signal transduction. Here, we modeled the azobenzenes as two-level systems, as shown in the inset to Fig. 4, where $n_g$ and $n_{ex}$ are the densities of dipoles in the ground (*trans*) and excited (*cis*) state, $\Gamma_2$ is the relaxation rate from the excited state to the ground state, and $\Gamma_1(I)$ is the transition rate from the ground state to the excited state and is directly proportional to power density $I$, $\Gamma_1(I)=\alpha I$, when the power is below the level at which nonlinear effects occur. Assuming that all molecules are in the ground state before illumination, the time dependence of $n_{ex}$ is given by

$$n_{ex}(t) = \frac{\Gamma_1}{\Gamma_1 + \Gamma_2} n_{tot} \left\{1 - \exp\left[-(\alpha I + \Gamma_2)t\right]\right\} \qquad (1)$$

where $n_{tot}$ is the sum of $n_g$ and $n_{ex}$, a constant value representing the density of azobenzenes around the SWNT in the device. Since the threshold voltage shift $\Delta V_g$ is proportional to $n_{ex}$, in the steady-state we have

$$\Delta V_g^{-1} = \left(\frac{n_{ex}}{n_{tot}}\frac{C_{NT}}{C_{tot}}\Delta V\right)^{-1} = \left(1+\frac{\Gamma_2}{\alpha}I^{-1}\right)\left(\frac{C_{NT}}{C_{tot}}\Delta V\right)^{-1} \qquad (2)$$

Here, $\Delta V$ is the difference in electrostatic potential on the SWNT between $n_g = n_{tot}$ and $n_{ex} = n_{tot}$. Thus, the inverse of the threshold voltage should depend linearly on the inverse of the illumination power density $I$. Furthermore, the inverse of the time constant for the change in conductance when the device is illuminated, $\tau^{-1} = \alpha I + \Gamma_2$, should depend linearly on the power density $I$.

To test this model, we measured the threshold voltage shift as a function of illumination intensity at a wavelength of 500 nm on devices functionalized with either DR1-PB or DO3-PB. The results are plotted in Fig. 4. The linear relationship between $\Delta V^{-1}$ and $I^{-1}$ agrees with the prediction of Eq. 2, implying that the observed transduction signals are indeed coming from a two-state behavior of the chromophores. We also functionalized devices with a 4:1 molar ratio hexyl-PB:DR1-PB solution. The dependence of threshold voltage shift of these devices on the power density still follows Eq. 2, but the magnitude of the shift is larger than ¼ of the shift seen on devices with pure DR1-PB. This effect could arise from preferable attachment of DR1-PB onto SWNTs over hexyl-PB. Next, time traces of devices with DR1-PB were taken at fixed wavelength (500 nm) but different power densities. The gate voltage was chosen to make the device conductance before illumination at 20-30% of its maximum post-illumination value, and was held constant during measurements. As Fig. 5a shows, the conductance of SWNT FETs increased upon illumination and slowly decayed back when light was turned off. A careful examination reveals that although the conductance initially increased following the form of Eq. 1, the rise slowed down after the first few seconds of illumination, as can be seen in Fig. 5b. In other words, there are two different time constants. The time constants are extracted from the slopes in Fig. 5c, where $1-I(t)/\Delta I_{tot} = 1-n_{ex}(t)/n_{tot}$ is plotted logarithmically against time with $\Delta I_{tot}$ being the total change in current induced by the illumination. As Fig. 5d shows, the fast time constant $\tau$ is about 1.5 s at 800 W/m$^2$, and $\tau^{-1}$ is directly proportional to the power density, in agreement with Eq. 1. On the other hand, the slow time constant, on the order of 20 s, does not depend on power density. Therefore, this suggests that the initial increase of SWNT conductance upon illumination is induced by the switching of molecular dipoles as the two-level system model in Eqs. 1 and 2 describes. Other light-insensitive dipoles around SWNTs, such as those of surface-bound water molecules that have been shown to dominate the hysteresis of SWNT FETs[19], may change their configurations, triggered by the switching of nearby chromophore dipoles.

Next, we analyzed the data on a more quantitative level and sought to extract intrinsic properties of these chromophores from our experiments. According to Eq. 1, the relationship between $\Gamma_1(I) = \alpha I$ and $\Gamma_2$ can be extracted from the linear fit in Fig. 4 since the number of molecules in the excited state is determined by the ratio of the two transition frequencies. For the device with DR1-PB, we find that *$\Gamma_2 = 240\alpha$*. The same relationship can be obtained from the kinetics of the signal as well, because $\Gamma_1$ and $\Gamma_2$ dictate the time it takes for the molecules to switch. The fit in Fig. 5d gives *$\Gamma_2 = 100\alpha$*. These two values obtained from different experiments, although different by a factor of 2.4, are reasonably close. In addition, we can deduce optical properties of the chromophores from our electrical measurements. For example, from the time constant $\tau$, the molecular absorptivity of DR1-PB was calculated to be $2\times10^3$ cm$^{-1}$M$^{-1}$. This number is about ten times smaller than the value measured from optical absorption studies[15]. This discrepancy is not surprising because the absorptivity is sensitive to the local environment. As previous studies have shown[15], absorptivity varies by more than 200% when chromophores are in different organic solvents. In our experiment, the chromophores are not in solvents, and are so closely packed that the dipole moments can couple together to reduce the transition frequency when photons are absorbed at low rates. We tried to solve this issue by diluting DR1-PB with light-insensitive, dipole-less hexyl-PB. The time constant $\tau$ was reduced slightly, but uncertainty involving the arrangement of these two types of molecules near SWNTs prevented us from drawing a solid conclusion and further work is underway.

Through both experimental and theoretical studies, this work showed that SWNTs can transduce the photoabsorption-induced isomerizations of nearby chromophores into electrical signals. By designing chromophores to absorb in a narrow window of the optical spectrum and applying them to SWNT FETs, sensitive nanoscale color detectors were demonstrated. This system can be used to study fundamental properties of chromophore-nanotube hybrids and to probe molecular transitions. We expect that improvements in the signal transduction will lead to the ability to detect single molecular transformation events, and that molecular engineering can provide detection in other regions of the optical spectrum.

**Acknowledgement.** The authors thank J.M. Simmons for valuable discussions. This project is supported by the Laboratory Directed Research and Development program at Sandia National Laboratories, a multiprogram laboratory operated by Sandia Corporation, a Lockheed Martin Company, for the United States Department of Energy under contract DE-AC04-94-AL85000.

**Supporting Information Available:** General procedure for chromophore-PB synthesis through esterification/amidation reaction of 1-pyrenebutyric acid with diazobenzene dyes, and characterizations of all three chromophores by $^1$H-NMR spectroscopy, FTIR, mass spectrometry, and UV/Vis spectrophotometry. This material is available free of charge via the Internet at http://pubs.acs.org.

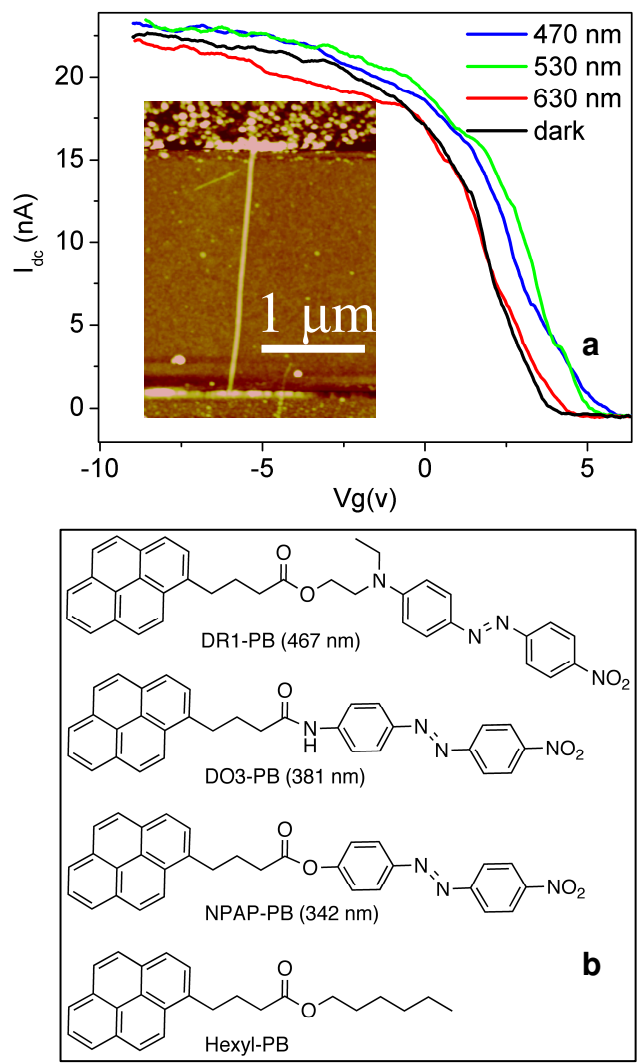

Figure 1. (a) Transfer characteristics of a DR1-PB functionalized SWNT-FET in the dark (black curve). The threshold voltage is shifted when the device is illuminated with either blue (blue curve) or green light (green curve), but remains constant when the device is illuminated with red light (red curve). Inset: AFM image of a device with a SWNT bridging a 2 µm gap. (b) Structures of pyrenebutyric acid functionalized chromophores. The numbers in parentheses are the absorbance maxima of the azobenzene moieties of the dyes.

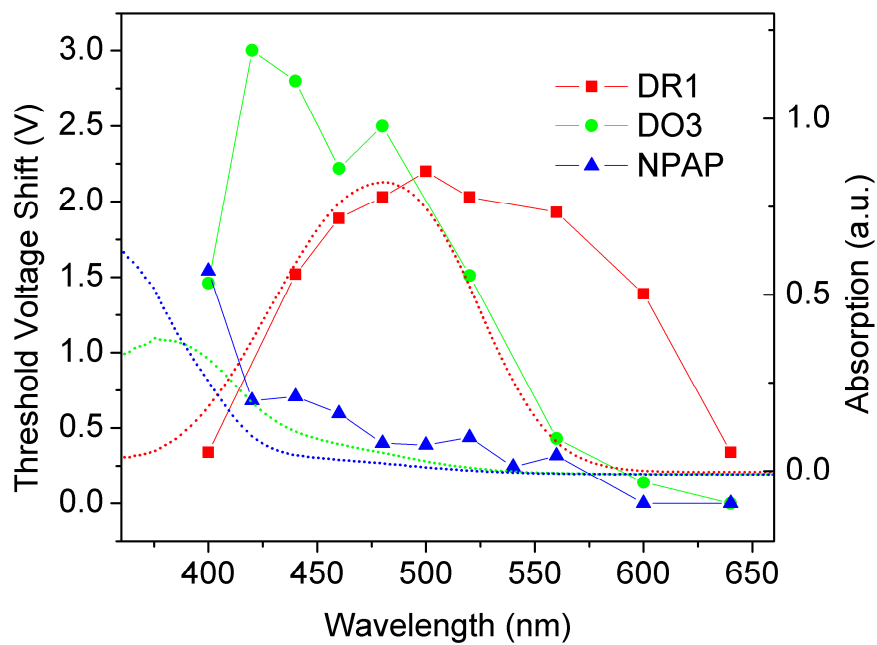

Figure 2. The threshold voltage shift of devices functionalized with different chromophores—DR1 (red squares), DO3 (green dots) and NPAP (blue triangles)—correlates with the absorption spectra (dotted curves) of these molecules.

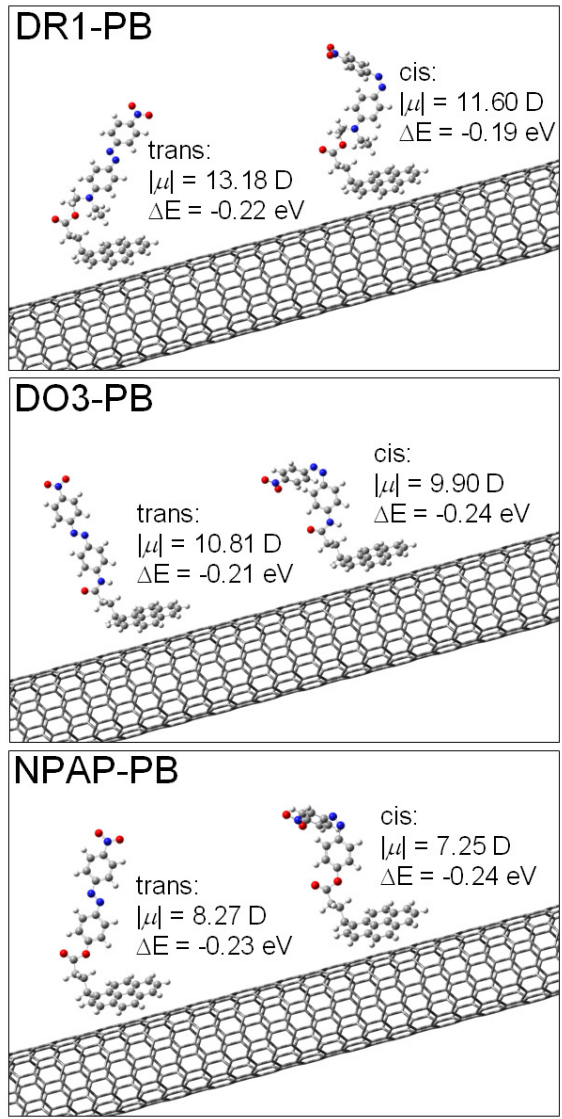

Figure 3. Optimized geometries, dipole moments, and binding energies for DR1-PB, DO3-PB, and NPAP-PB noncovalently attached to a (10,0) nanotube. Each of the pyrene units is in contact with the nanotube surface at a distance of about 2.8 Å.

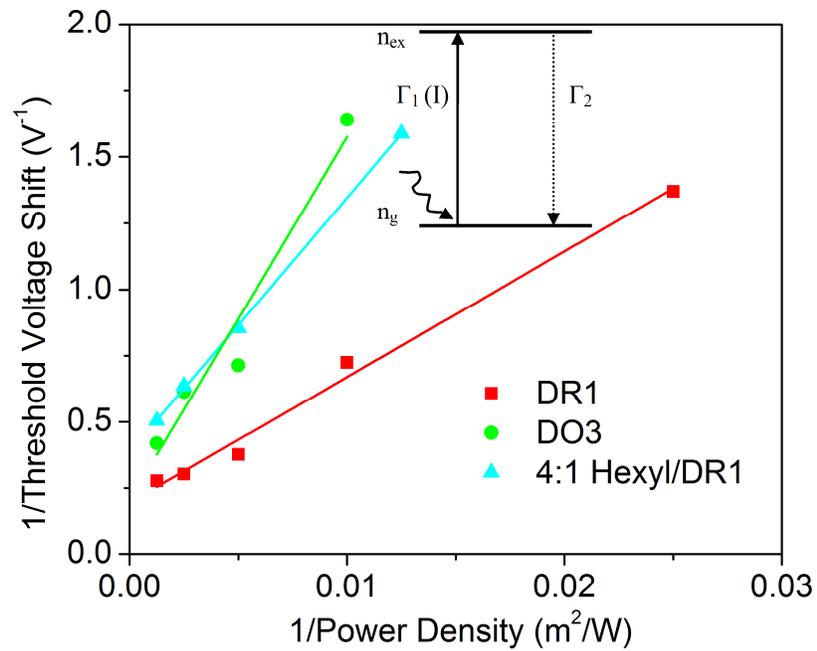

Figure 4. The inverse of the threshold voltage shift depends linearly on the inverse of the illumination power density. Solid lines are linear fits to the data for DR1-PB (red squares), DO3-PB (green dots) and 4:1 hexyl-PB/DR1-PB (cyan triangles). Inset: Schematic of the two-level system used to model the experiments.

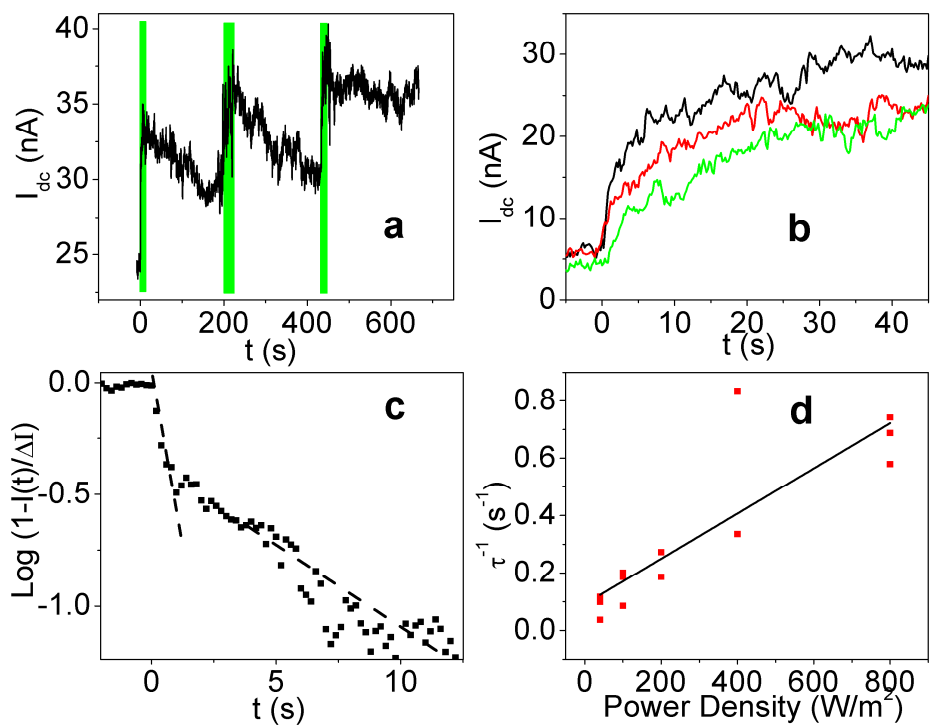

Figure 5. Temporal response of chromophore-functionalized devices upon illumination. (a) The conduction of a DR1-PB functionalized device increases rapidly upon illumination and drops back slowly when the illumination is turned off. (b) The conduction rise of the same device for different intensities: 200 (black), 100 (red), and 40 (green) W/m$^2$. (c) When the change in conduction is plotted logarithmically against time, two time scales can be seen. The fast one ($\tau \sim 2$ s) dominates for the first few seconds after illumination is turned on, but the change slows down afterwards. (d) The inverse of the fast time scale $\tau$ is directly proportional to the power density. $\tau$ of the same device is measured three times at each power density and all three values are shown in the plot.

Supporting Information

**General procedure for esterification/amidation reaction of 1-pyrenebutyric acid with diazobenzene dyes**:

The dye, 1-pyrenebutyric acid, dicyclohexylcarbodiimide (DCC), and 4-(dimethylamino)pyridine (dmap) were dissolved in anhydrous dichloromethane with stirring. The reaction mixture was stirred until complete (up to 18 hours). The precipitated dicyclohexyl urea was removed by vacuum filtration and the crude product isolated by rotary evaporation of the solvent. Products were purified by recrystallization or column chromatography (silica gel).

**(*E*)-2-(ethyl(4-((4-nitrophenyl)diazenyl)phenyl)amino)ethyl 4-(pyren-1-yl)butanoate (DR1-PB)**: Using the general procedure above, disperse red 1 (500 mg, 1.59 mmol), 1-pyrenebutryic acid (462 mg, 1.60 mmol), DCC (361 mg, 1.75 mmol) and DMAP (39 mg, 0.32 mmol) were stirred in anhydrous dichloromethane for 18 hours. After filtration of the precipitate and collection of the crude product by rotary evaporation, the red powder was recrystallized from ethyl acetate/ethanol to give a dark red powder (667 mg, 72%). $^1$H-NMR (500 MHz, CDCl$_3$): δ 8.30 (d, 2H), 8.26 (d, 1H), 8.15 (d, 2H), 8.09 (m, 2H), 8.02-7.92 (m, 7H), 7.82 (d, 1H), 6.80 (d, 2H), 4.28 (t, 2H), 3.63 (t, 2H), 3.49 (q, 2H), 3.36 (t, 2H), 2.45 (t, 2H), 2.20 (m, 2H), 1.21 (t, 3H); UV/Vis: azobenzene $\lambda_{max}$ 467 nm; FTIR: C=O stretch 1745 cm$^{-1}$; MS (APCI, positive ion mode) Calcd for C$_{36}$H$_{32}$N$_4$O$_4$: 584.2, Found: 585.4 [M+H]$^+$.

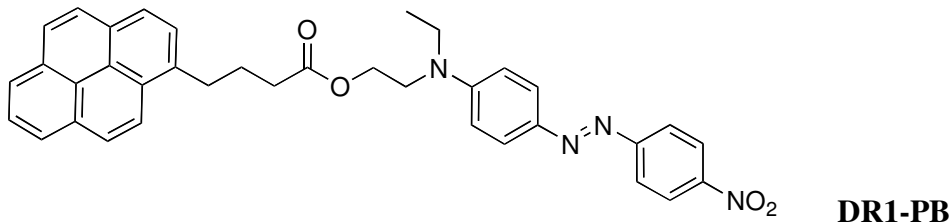

**DR1-PB**

(*E*)-*N*-(4-((4-nitrophenyl)diazenyl)phenyl)-4-(pyren-1-yl)butanamide (**DO3-PB**): Using the general procedure above, disperse orange 3 (480 mg, 2.0 mmol), 1-pyrenebutryic acid (650 mg, 2.25 mmol), DCC (610 mg, 3.0 mmol) and DMAP (30 mg, 0.2 mmol) were stirred in anhydrous dichloromethane for 18 hours. After filtration of the precipitate and collection of the crude product by rotary evaporation, the orange-red powder was recrystallized from ethyl acetate/ethanol to give a red powder (635 mg, 63 %). $^1$H-NMR (500 MHz, DMSO-d$_6$): δ 8.42 (d, 2H), 8.35 (d, 2H), 8.13 (m, 2H), 7.88 (m, 7H), 7.74 (d, 2H), 6.70 (d, 2H), 6.50 (bs, 1H), 3.26 (t, 2H), 2.62 (t, 2H), 1.22 (m, 2H); UV/Vis: azobenzene λ$_{max}$ 381 nm; FTIR: C=O stretch 1692 cm$^{-1}$; MS (APCI, negative ion mode) Calcd for C$_{32}$H$_{24}$N$_4$O$_3$: 512.2, Found: 512.2.

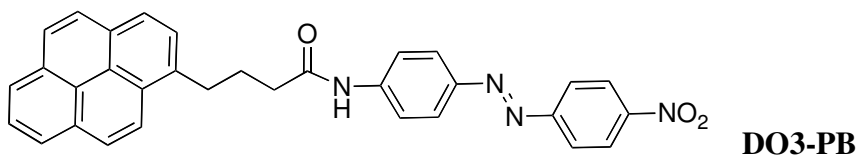

**DO3-PB**

(*E*)-4-((4-nitrophenyl)diazenyl)phenyl 4-(pyren-1-yl)butanoate (**NPAP-PB**): Using the general procedure above, 4-(4-nitrophenyl)azophenol (490 mg, 2.0 mmol), 1-pyrenebutryic acid (650 mg, 2.25mmol), DCC (540 mg, 2.6 mmol) and DMAP (20 mg, 0.16 mmol) were stirred in anhydrous dichloromethane for 18 hours. After filtration of the precipitate and collection of the crude product by rotary evaporation, the dark red powder was purified on a silica column from ethyl acetate/petroleum ether (50/50) to give a dark red powder (138 mg, 13%). $^1$H-NMR (500 MHz, DMSO-$d_6$): δ 8.45 (d, 2H), 8.38 (d, 2H), 8.26 (m, 5H), 8.14 (d, 2H), 7.87 (d, 2H), 7.40 (d, 2H), 6.97 (d, 2H), 3.46 (t, 2H), 2.81 (t, 2H), 2.18 (m, 2H); UV/Vis: azobenzene $\lambda_{max}$ 342 nm; FTIR: C=O stretch 1763 cm$^{-1}$; MS (APCI, negative ion mode) Calcd for $C_{32}H_{23}N_3O_4$: 513.2, Found: 513.2.

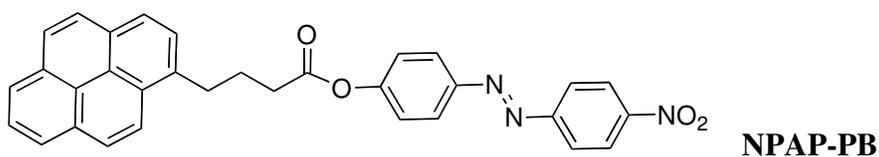

**NPAP-PB**

**Hexyl 4-(pyren-1-yl)butanoate (hexyl-PB)**: Using the general procedure above, 1-hexanol (205 mg, 2.0mmol), 1-pyrenebutryic acid (470 mg, 1.6mmol), DCC (390 mg, 1.8mmol) and DMAP (40 mg, 0.3mmol) were stirred in anhydrous dichloromethane for 18 hours. After filtration of the precipitate and collection of the oily crude product by rotary evaporation, the product was purified on a silica gel column with ethyl acetate/petroleum ether (30/70) to give a yellow oil (596 mg, 59%). $^1$H-NMR (500 MHz, DMSO-$d_6$): δ 8.37 (d, 1H), 8.26 (t, 2H), 8.22 (m, 2H), 8.12, (d, 2H), 8.05 (t, 1H), 7.93 (d, 1H), 3.98 (t, 2H), 3.33 (t, 2H), 2.45 (t, 2H), 2.02 (t, 2H), 1.51 (t, 2H), 1.22 (m, 6H), 0.81 (t, 3H); FTIR: C=O stretch 1728 cm$^{-1}$; MS (APCI, positive ion mode) Calcd for $C_{26}H_{28}O_2$: 372.2, Found: 373.3 [M+H]$^+$.

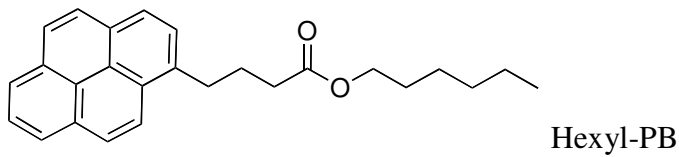

Hexyl-PB